\documentclass[aps,prl,twocolumn,superscriptaddress,showpacs]{revtex4-2}
\usepackage[colorlinks,citecolor=blue,urlcolor=blue,linkcolor=blue,bookmarks=false,hypertexnames=true]{hyperref}
\usepackage{xcolor}
\usepackage{multirow}

\usepackage{graphicx}  
\usepackage{dcolumn}   
\usepackage{bm}        
\usepackage{amssymb}   
\usepackage{amsmath}   
\usepackage{ulem}

\begin{document}

\title{
   Observation of a broad $p$-wave resonant state in $^{9}$He
}

\newcommand{\abuaa}{        \affiliation{School of Physics, Beihang University, Beijing 100191, China}}
\newcommand{\akth}{        \affiliation{Department of Physics, Royal Institute of Technology, SE-10691 Stockholm, Sweden}}
\newcommand{\aatomki}{     \affiliation{MTA Atomki, P.O. Box 51, Debrecen H-4001, Hungary}}
\newcommand{\abeijing}{    \affiliation{State Key Laboratory of Nuclear Physics and Technology, Peking University, Beijing 100871, P.R. China}}
\newcommand{\acaen}{       \affiliation{Université de Caen Normandie, ENSICAEN, CNRS/IN2P3, LPC-Caen UMR6534, F-14000 Caen, France}}
\newcommand{\acea}{        \affiliation{IRFU, CEA, Universit\'e Paris-Saclay, F-91191 Gif-sur-Yvette, France}}
\newcommand{\acns}{        \affiliation{Center for Nuclear Study, University of Tokyo, RIKEN campus, Wako, Saitama 351-0198, Japan}}
\newcommand{\aewha}{       \affiliation{Department of Physics, Ewha Womans University, Seoul, South Korea}}
\newcommand{\agsi}{        \affiliation{GSI Helmoltzzentrum f\"ur Schwerionenforschung GmbH, Planckstr. 1, 64291 Darmstadt, Germany}}
\newcommand{\ahku}{        \affiliation{Department of Physics, The University of Hong Kong, Pokfulam, Hong Kong}}
\newcommand{\ainst}{       \affiliation{Institute for Nuclear Science \& Technology, VINATOM, 179 Hoang Quoc Viet, Cau Giay, Hanoi, Vietnam}}
\newcommand{\aipno}{       \affiliation{Institut de Physique Nucl\'eaire Orsay, IN2P3-CNRS, F-91406 Orsay Cedex, France}}
\newcommand{\akoeln}{      \affiliation{Institut f\"ur Kernphysik, Universit\"at zu K\"oln, D-50937 Cologne, Germany}}
\newcommand{\alanzhou}{    \affiliation{Institute of Modern Physics, Chinese Academy of Sciences, Lanzhou, China}}
\newcommand{\amadrid}{     \affiliation{Instituto de Estructura de la Materia, CSIC, E-28006 Madrid, Spain}}
\newcommand{\aorsay}{      \affiliation{CSNSM, CNRS/IN2P3, Universit\'e Paris-Sud, F-91405 Orsay Campus, France}}
\newcommand{\aoslo}{       \affiliation{Department of Physics, University of Oslo, N-0316 Oslo, Norway}}
\newcommand{\ariken}{      \affiliation{RIKEN Nishina Center, 2-1 Hirosawa, Wako, Saitama 351-0198, Japan}}
\newcommand{\arikkyo}{     \affiliation{Department of Physics, Rikkyo University, 3-34-1 Nishi-Ikebukuro, Toshima, Tokyo 172-8501, Japan}}
\newcommand{\atitech}{     \affiliation{Department of Physics, Institute of Science Tokyo, 2-12-1 O-Okayama, Meguro, Tokyo, 152-8550, Japan}}
\newcommand{\atohoku}{     \affiliation{Department of Physics, Tohoku University, Sendai 980-8578, Japan}}
\newcommand{\atud}{        \affiliation{Institut f\"ur Kernphysik, Technische Universit\"at Darmstadt, 64289 Darmstadt, Germany}}
\newcommand{\aunal}{       \affiliation{Universidad Nacional de Colombia, Carr. 30 No. 45-03, Bogot\'a, Colombia}}
\newcommand{\atokyo}{      \affiliation{Department of Physics, University of Tokyo, 7-3-1 Hongo, Bunkyo, Tokyo 113-0033, Japan}}
\newcommand{\aoak}{        \affiliation{Physics Division, Oak Ridge National Laboratory, Oak Ridge, Tennessee 37831, USA}}
\newcommand{\atennessee}{  \affiliation{Department of Physics and Astronomy, University of Tennessee, Knoxville, Tennessee 37996, USA}}
\newcommand{\atriumf}{     \affiliation{TRIUMF 4004 Wesbrook Mall, Vancouver, British Columbia V6T 2A3, Canada}}
\newcommand{\amaxplank}{   \affiliation{Max-Planck-Institut f\"ur Kernphysik, Saupfercheckweg 1, 69117 Heidelberg, Germany}}
\newcommand{\arcnp}{       \affiliation{Research Center for Nuclear Physics (RCNP), Osaka University, Ibaraki 567-0047, Japan}}
\newcommand{\ajaea}{       \affiliation{Japan Atomic Energy Agency, Tokai, Ibaraki 319-1195, Japan}}
\newcommand{\asurrey}{     \affiliation{Department of Physics, University of Surrey, Guildford GU2 7XH, UK}}
\newcommand{\aku}{         \affiliation{KU Leuven, Instituut voor Kern- en Stralingsfysica, 3001 Leuven, Belgium}}
\newcommand{\astrasbourg}{ \affiliation{Universit\' de Strasbourg, IPHC, 67037 Strasbourg Cedex, France}}
\newcommand{\atokuyama}{   \affiliation{National Institute of Technology, Tokuyama College, Shunan 745-8585, Japan}}
\newcommand{\apku}{        \affiliation{State Key Laboratory of Nuclear Physics and Technology, School of Physics, Peking University, Beijing 100871, China}}
\newcommand{\amiyazaki}{   \affiliation{Department of Applied Physics, University of Miyazaki, Gakuen-Kibanadai-Nishi 1-1, Miyazaki 889-2192, Japan}}
\newcommand{\akyushu}{     \affiliation{Department of Physics, Kyushu University, Fukuoka 812-8581, Japan}}
\newcommand{\atum}{        \affiliation{Department of Physics, Technische Universit\"at M\"unchen, James-Franck-Stra{\ss}e 1, 85748 Garching, Germany}}
\newcommand{\akyoto}{      \affiliation{Department of Physics, Kyoto University, Kitashirakawa, Sakyo, Kyoto 606-8502, Japan}}
\newcommand{\ausa}{      \affiliation{Departamento de F\'isica At\'omica, Molecular y Nuclear, Facultad de F\'isica, Universidad de Sevilla, Apartado 1065, E-41080 Sevilla, Spain}}
\newcommand{\atexas}{   \affiliation{Department of Physics \& Astronomy and Cyclotron Institute, Texas A\&M University, College Station, TX 77843, USA}}
\newcommand{\auec}{   \affiliation{Department of Engineering Science, University of Electro-Communications, Chofu, Tokyo 182-8585, Japan}}
\newcommand{\ayork}{  \affiliation{School of Physics, Engineering and Technology, University of York, York YO10 5DD, United Kingdom}}

\author{Y.~L.~Sun} \email{sunyelei@buaa.edu.cn} \abuaa \atud \acea
\author{A.~Corsi} \acea
\author{Y.~Kubota} \ariken \acns \atud
\author{G.~Authelet} \acea
\author{H.~Baba} \ariken
\author{C.~Caesar} \atud
\author{D.~Calvet} \acea
\author{A.~Delbart} \acea
\author{M.~Dozono} \acns
\author{J.~Feng} \apku
\author{F.~Flavigny} \aipno
\author{J.-M.~Gheller} \acea
\author{J.~Gibelin} \acaen
\author{A.~Giganon} \acea
\author{A.~Gillibert} \acea
\author{S.~Giraud} \acea
\author{K.~Hasegawa} \atohoku
\author{T.~Isobe} \ariken
\author{Y.~Kanaya} \amiyazaki
\author{S.~Kawakami} \amiyazaki
\author{D.~Kim} \aewha
\author{Y.~Kiyokawa}\acns
\author{M.~Kobayashi}\acns
\author{N.~Kobayashi} \atokyo
\author{T.~Kobayashi} \atohoku
\author{Y.~Kondo} \atitech
\author{Z.~Korkulu} \ariken
\author{S.~Koyama} \atokyo
\author{V.~Lapoux} \acea
\author{Y.~Maeda} \amiyazaki
\author{F.~M.~Marqu\'{e}s} \acaen
\author{T.~Miyazaki} \atokyo
\author{T.~Motobayashi} \ariken
\author{T.~Nakamura} \atitech
\author{N.~Nakatsuka} \akyoto
\author{Y.~Nishio} \akyushu
\author{A.~Obertelli} \atud \acea
\author{A.~Ohkura} \akyushu
\author{N.~A.~Orr} \acaen
\author{S.~Ota} \acns
\author{H.~Otsu} \ariken
\author{T.~Ozaki} \atokyo
\author{V.~Panin} \ariken \acea
\author{S.~Paschalis} \atud \ayork
\author{E.~C.~Pollacco} \acea
\author{S.~Reichert} \atum
\author{J.-Y.~Rouss\'e} \acea
\author{A.~T.~Saito} \atitech
\author{S.~Sakaguchi} \akyushu
\author{M.~Sako} \ariken
\author{C.~Santamaria} \acea
\author{M.~Sasano} \ariken
\author{H.~Sato} \ariken
\author{M.~Shikata} \atitech
\author{Y.~Shimizu} \ariken
\author{Y.~Shindo} \akyushu
\author{L.~Stuhl} \ariken
\author{T.~Sumikama} \ariken
\author{M.~Tabata} \akyushu
\author{Y.~Togano} \atitech
\author{J.~Tsubota} \atitech
\author{Z.~H.~Yang} \ariken
\author{J.~Yasuda} \akyushu
\author{K.~Yoneda} \ariken
\author{J.~Zenihiro} \ariken
\author{T.~Uesaka} \ariken

\begin{abstract}
We report on the two-body invariant-mass spectroscopy of $^{9}$He,
populated via the 1$p$1$n$ knockout reaction from the two-neutron halo nucleus $^{11}$Li
at $\sim$~250~MeV/nucleon.
A broad $p$-wave resonant state
of $^{9}$He was observed at 1.28(1)~MeV with a width of 0.82(4)~MeV.

\end{abstract}


\date{\today}
\maketitle


The $^{9}$He nucleus, consisting of an $^{8}$He core and one valence neutron has been extensively studied.
The ground state of $^{9}$He was first found unbound against single-neutron decay by 1.13(10)~MeV~\cite{Seth1987} and 1.27(10)~MeV~\cite{Oertzen1995,Bohlen1999} 
with a width of about 100 keV, and was first interpreted as a $\textit{p}$-wave resonance ($\textit{J$^{\pi}$}$ = 1/2$^{-}$)~\cite{Oertzen1995,Bohlen1999}.
The study of isobaric analog states of $^{9}$He in $^{9}$Li suggests a narrow $p$-wave resonance at 1.1 MeV~\cite{Rogachev2003}.
Later, the $\textit{p}$-wave resonance
was identified
at 2.0(2)~MeV with $\Gamma$ $\sim$~2~MeV
and at 1.2(1)~MeV with $\Gamma$ of 130$^{+170}_{-130}$~keV in 
the $\textit{d}$($^{8}$He, $\textit{p}$)$^{9}$He transfer reactions with poor statistics~\cite{Golovkov2007, Kalanee2013}.
In addition to the $\textit{p}$-wave resonance, the $^{9}$He spectrum also shows an enhancement close to the $^{8}$He-$\textit{n}$ decay threshold, 
which is interpreted as an $\textit{s}$-wave virtual state ($\textit{J$^{\pi}$}$ = 1/2$^{+}$). 
So far, the experimentally extracted $s$-wave scattering lengths that quantifies the $^{9}$He low-energy virtual state are rather divergent.
An upper limit of $a_{s}$ $\leq$ $-$~10~fm was determined using the $^{11}$Be(-2$p$) knockout reaction~\cite{Chen2001},
while a range of $-$3~fm $\textless$ $a_{s}$ $\textless$ 0~fm was suggested using the same reaction~\cite{Falou2011}.
Recently, two possibilities around $-$2~fm and $-$7~fm were also obtained~\cite{Votaw2020}.
Meanwhile, the $^{11}$Li(-1$p$1$n$) knockout reaction suggested $a_{s}$ = $-$3.17(66)~fm~\cite{Johansson2010}.
A lower limit of $a_{s}$ $\textgreater$ $-$20~fm~\cite{Golovkov2007}
and a value of $a_{s}$ = $-$12(3)~fm~\cite{Kalanee2013}
were obtained from the $^{8}$He induced transfer reactions.
Finally, the most recent high resolution $^{8}$He~+~$\textit{p}$ elastic scattering that populates isobaric 
analogue $^{9}$He ground and excited states 
showed that
$^{9}$He does not have narrow resonances between 0 and 2.2~MeV and 
its lowest state is a broad
$\textit{s}$-wave resonance around 3~MeV which is most probably a virtual state~\cite{Uberseder2016}.
A limit of $-$1.7~fm $\textless$ $a_{s}$ $\textless$ 0~fm, $\textit{i.e.}$ close to zero,
was suggested from the phase shift analysis~\cite{Uberseder2016,Rogachev_private}.
These results suggest that the $^{8}$He-$\textit{n}$ $s$-wave interaction could be either strong ($a_{s}$ $\leq$ $-$10~fm) or weak ($a_{s}$ close to zero),
making it difficult to conclude on the $^{9}$He structure 
as well as the influence on the $^{10}$He structure~\cite{Grigorenko2008}.
Further experimental and theoretical studies are necessary to understand the structure of $^{9}$He and $^{10}$He consistently.

   \begin{figure}[t!]
   \setlength{\abovecaptionskip}{0pt}
   \setlength{\belowcaptionskip}{0pt}
   \centering
   \includegraphics[width=8.5cm]{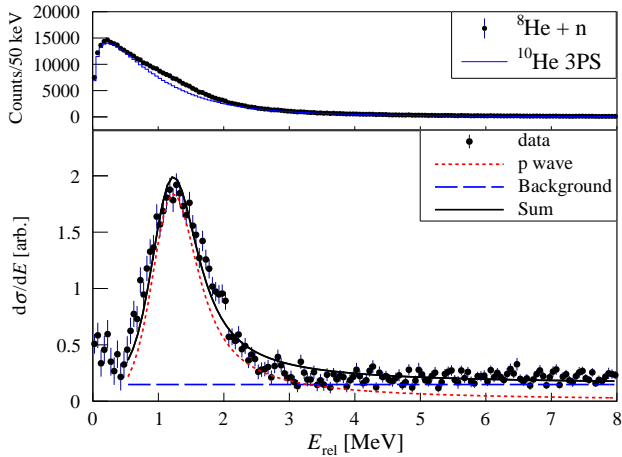}
      \caption{
      (Color online) (top panel) Experimental $^{8}$He-$\textit{n}$ relative energy spectrum obtained 
      from the $^{11}$Li(-1$\textit{p}$1$\textit{n}$) reaction.
      Contaminations from the $^{10}$He three-body phasespace decay (3PS) is shown by the blue-solid line.
      (lower panel) Spectrum after the contaminations subtraction and efficiency correction. 
      The black solid line shows the fit results of a Breit-Wigner shape $p$-wave resonance folded with the experimental resolution (red short dashed line) on top of a constant background (blue long dashed line).
   }
   \label{fig:9He}
   \end{figure}

    Here, we report on the low-lying state of $^{9}$He populated from the $^{11}$Li(-1$p$1$n$) knockout reaction at $\sim$~250~MeV/nucleon.

   The experiment was performed at the Radioactive Isotope Beam Factory operated by the RIKEN Nishina Center 
and the Center for Nuclear Study of the University of Tokyo.
   The cocktail secondary beam was produced through fragmentation of a 345~MeV/nucleon $^{48}$Ca primary beam on a $^{9}$Be target.
   The typical intensity of the primary beam is 400 particle nA.
   The secondary beam was
   purified and identified event by event using the BigRIPS two-stage fragment separator~\cite{Kubo2012, Fukuda2013}.
   The $^{11}$Li beam, with an average energy of 246~MeV/nucleon and a typical intensity of 1 $\times$ 10$^{5}$ particles per second, 
   was tracked by two multiwire drift chambers,
   and bombarded on the 150-mm thick liquid hydrogen target~\cite{Louchart2014} 
   of the MINOS device ~\cite{Obertelli2014} to induce one-nucleon or two-nucleon knockout reactions.
   The $^{8}$He fragment and decay neutrons were detected by the SAMURAI spectrometer~\cite{Kobayashi2013} and the neutron detector array NEBULA~\cite{Kondo2020}, respectively.
   The same setup has been used in previous publications~\cite{Corsi2019, Kubota2020, Yang2021, Monteagudo2024, Andre2024}.

   The two-body relative energy $\textit{E}\rm_{rel}$, which is also the energy of $^{9}$He above the $^{8}$He + $n$ threshold, 
   was reconstructed by the momenta of $^{8}$He and the neutron,
   requiring that only one neutron was detected in the neutron detector array.
   In the present work, $^{9}$He can be populated via $^{11}$Li($\textit{p}$,$\textit{pd}$) or $^{11}$Li($\textit{p}$,2$\textit{pn}$) reactions.
   $^{8}$He + $n$ events can also be produced if $^{10}$He is populated in the $^{11}$Li($\textit{p}$,2$\textit{p}$)$^{10}$He reaction and decays to $^{8}$He + 2$n$, but only one neutron was detected in the neutron detectors~\cite{Korsheninnikov1994}.
   Such contaminations were estimated using Monte Carlo simulations based on the quasi-free one-proton knockout process from $^{11}$Li,
   following by the three-body phasespace decay of $^{10}$He to $^{8}$He + 2$\textit{n}$ (3PS).
   The experimental energy distribution of $^{10}$He has been adopted as an input for the phasespace decay.
   Normalization was performed such that the two-neutron event number in the simulation matches the corresponding experimental data.
   The simulations reproduce well the experimental relative-energy spectrum of the $^{8}$He-$\textit{n}$ subsystem from the $^{10}$He decay.

   The obtained $^{9}$He $\textit{E}\rm_{rel}$ spectrum as well as the contaminations from the $^{10}$He 3PS are shown in the top panel of Fig.~\ref{fig:9He}.
   After the contaminations subtraction and efficiency correction, 
   a clear peak was observed in the $\textit{E}\rm_{rel}$ spectrum, 
   as shown in the lower panel of Fig.~\ref{fig:9He},
   which is different with the previous spectrum measured at GSI at similar incident beam energy~\cite{Johansson2010}.

   \begin{figure}[t!]
   \setlength{\abovecaptionskip}{0pt}
   \setlength{\belowcaptionskip}{0pt}
   \centering
   \includegraphics[width=8.5cm]{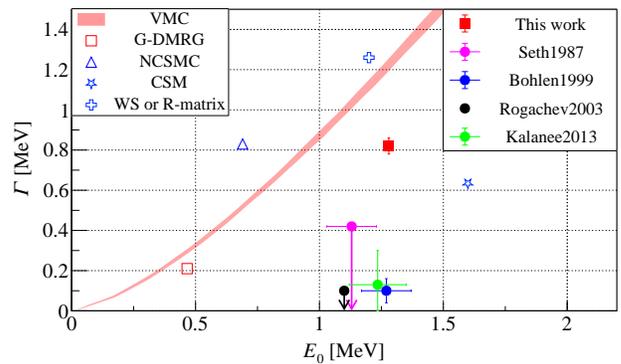}
      \caption{
      (Color online)
      The two-body $p$-wave resonant energy and decay width of $^{9}$He obtained by the present work compared with previous results from Seth $\textit{et al.}$~\cite{Seth1987}, Bohlen $\textit{et al.}$~\cite{Bohlen1999}, Rogachev $\textit{et al.}$~\cite{Rogachev2003} and Kalanee $\textit{et al.}$~\cite{Kalanee2013}. The experimental data are also compared with the different theoretical predictions, including $\textit{ab initio}$ variational Monte Carlo (VMC)~\cite{Nollett2012}, Gamow-density-matrix renormalization-group (G-DMRG)~\cite{Fossez2018}, no-core shell model with continuum (NCSMC)~\cite{Vorabbi2018}, the shell model in continuum (CSM)\cite{Volya2005} as well as the Woods–Saxon (WS) potential and the R-matrix analysis of Ref.~\cite{Barker2004}.
   }
   \label{fig2}
   \end{figure}

   The $\textit{E}\rm_{rel}$ spectrum of $^{9}$He was fitted with a Breit-Wigner $p$-wave resonance on top of a constant background,
   convoluted with the experimental relative-energy resolution.
   The energy resolution (FWHM) was obtained from GEANT4 simulations that incorporated the realistic experimental setup and detector response,
   yielding a value of 0.40 MeV at 1.3 MeV.
   Both the peak position and decay width of the resonance were taken as free parameters in the fitting.
   Energy dependence of the decay width has been taken into account~\cite{Aksyutina2008,Lane1958}.
   The resulting resonant energy and decay width are $E_{0}$ = 1.28(1)~MeV and $\Gamma$ = 0.82(4)~MeV, respectively.
   The quoted uncertainties are only statistical.
   In addition, the channel radius is set as 4.35~fm in the fitting~\cite{Barker2004}.
   Increasing the channel radius to 6.0~fm reduces the obtained decay width by about 4\%.

   The obtained $p$-wave resonant energy of $^{9}$He is 
   $\sim$0.7 MeV lower than that observed by $^{8}$He($d$,$p$) transfer reaction~\cite{Golovkov2007}, but
   shows consistency with results
   from the double-charge exchange reaction~\cite{Seth1987}, other transfer reactions~\cite{Oertzen1995,Bohlen1999,Kalanee2013} and ($p$,$p$) resonance elastic scattering~\cite{Rogachev2003}.
   However, the measured decay width is nearly an order of magnitude larger than the previously reported values --- 0.10(6)~MeV~\cite{Bohlen1999}, $\textless$ 0.1 MeV~\cite{Rogachev2003} and 130$^{+170}_{-130}$~keV~\cite{Kalanee2013} --- all of which were 
   derived from datasets with limited statistical significance.

   On the other hand, the current data can also serve as a benchmark for various $\textit{ab initio}$ and phenomenological nuclear models.
   The present data, in particularly the much larger decay width around 1 MeV, agrees reasonably well with the shell model in continuum (CSM) calculations~\cite{Volya2005}, the
   $\textit{ab initio}$ variational Monte Carlo (VMC) calculations~\cite{Nollett2012} and the no-core shell model with continuum (NCSMC) predictions~\cite{Vorabbi2018}, as well as the
   Woods–Saxon (WS) potential and the R-matrix analysis of Ref.~\cite{Barker2004}.
   These theoretical models anticipate a decay width around ~1~MeV for the lowest 1/2$^{-}$ state in $^{9}$He,
   that can be attributed to its strong single-particle character.
   Notably, due to the deuteron spin and parity of 1$^{+}$, $^{9}$He populated from the $^{11}$Li($\textit{p}$,$\textit{pd}$) reaction
   can only
   exhibit a 1/2$^{-}$ configuration
   ($\textit{i.e.}$ $^{8}$He $\otimes$ $\nu$0\textit{p}$_{1/2}$).
   This selection rule likely explains the absence or weak population of the 1/2$^{+}$ state of $^{9}$He (Fig.~\ref{fig:9He} below 0.5~MeV) in the present energy spectrum.

\section*{References}
\bibliography{mybibfile}

@article{Korsheninnikov1994,
title = "Observation of 10He",
journal = "Physics Letters B",
volume = "326",
number = "1",
pages = "31 - 36",
year = "1994",
issn = "0370-2693",
doi = "https://doi.org/10.1016/0370-2693(94)91188-6",
url = "http://www.sciencedirect.com/science/article/pii/0370269394911886",
author = "A.A. Korsheninnikov and K. Yoshida and D.V. Aleksandrov and N. Aoi and Y. Doki and N. Inabe and M. Fujimaki and T. Kobayashi and H. Kumagai and C.-B. Moon and E.Yu. Nikolskii and M.M. Obuti and A.A. Ogloblin and A. Ozawa and S. Shimoura and T. Suzuki and I. Tanihata and Y. Watanabe and M. Yanokura"
}

@article{Johansson2010,
title = "The unbound isotopes 9,10He",
journal = "Nuclear Physics A",
volume = "842",
number = "1",
pages = "15 - 32",
year = "2010",
issn = "0375-9474",
doi = "https://doi.org/10.1016/j.nuclphysa.2010.04.006",
url = "http://www.sciencedirect.com/science/article/pii/S0375947410004057",
author = "H.T. Johansson and Yu. Aksyutina and T. Aumann and K. Boretzky and M.J.G. Borge and A. Chatillon and L.V. Chulkov and D. Cortina-Gil and U. Datta Pramanik and H. Emling and C. Forssén and H.O.U. Fynbo and H. Geissel and G. Ickert and B. Jonson and R. Kulessa and C. Langer and M. Lantz and T. LeBleis and K. Mahata and M. Meister and G. Münzenberg and T. Nilsson and G. Nyman and R. Palit and S. Paschalis and W. Prokopowicz and R. Reifarth and A. Richter and K. Riisager and G. Schrieder and H. Simon and K. Sümmerer and O. Tengblad and H. Weick and M.V. Zhukov"
}

@article{Grigorenko2008,
  title = {Problems with the interpretation of the $^{10}\mathrm{He}$ ground state},
  author = {Grigorenko, L. V. and Zhukov, M. V.},
  journal = {Phys. Rev. C},
  volume = {77},
  issue = {3},
  pages = {034611},
  numpages = {13},
  year = {2008},
  month = {Mar},
  publisher = {American Physical Society},
  doi = {10.1103/PhysRevC.77.034611},
  url = {https://link.aps.org/doi/10.1103/PhysRevC.77.034611}
}

@article{Golovkov2007,
  title = {New insight into the low-energy $^{9}\mathrm{He}$ spectrum},
  author = {Golovkov, M. S. and Grigorenko, L. V. and Fomichev, A. S. and Gorshkov, A. V. and Gorshkov, V. A. and Krupko, S. A. and Oganessian, Yu. Ts. and Rodin, A. M. and Sidorchuk, S. I. and Slepnev, R. S. and Stepantsov, S. V. and Ter-Akopian, G. M. and Wolski, R. and Korsheninnikov, A. A. and Nikolskii, E. Yu. and Kuzmin, V. A. and Novatskii, B. G. and Stepanov, D. N. and Roussel-Chomaz, P. and Mittig, W.},
  journal = {Phys. Rev. C},
  volume = {76},
  issue = {2},
  pages = {021605},
  numpages = {5},
  year = {2007},
  month = {Aug},
  publisher = {American Physical Society},
  doi = {10.1103/PhysRevC.76.021605},
  url = {https://link.aps.org/doi/10.1103/PhysRevC.76.021605}
}

@article{Seth1987,
  title = {Exotic Nucleus Helium-9 and its Excited States},
  author = {Seth, Kamal K. and Artuso, M. and Barlow, D. and Iversen, S. and Kaletka, M. and Nann, H. and Parker, B. and Soundranayagam, R.},
  journal = {Phys. Rev. Lett.},
  volume = {58},
  issue = {19},
  pages = {1930--1933},
  numpages = {0},
  year = {1987},
  month = {May},
  publisher = {American Physical Society},
  doi = {10.1103/PhysRevLett.58.1930},
  url = {https://link.aps.org/doi/10.1103/PhysRevLett.58.1930}
}

@article{Chen2001,
title = "Evidence for an l=0 ground state in 9He",
journal = "Physics Letters B",
volume = "505",
number = "1",
pages = "21 - 26",
year = "2001",
issn = "0370-2693",
doi = "https://doi.org/10.1016/S0370-2693(01)00313-6",
url = "http://www.sciencedirect.com/science/article/pii/S0370269301003136",
author = "L Chen and B Blank and B.A Brown and M Chartier and A Galonsky and P.G Hansen and M Thoennessen"
}

@article{Kubo2012,
    author = {Kubo, Toshiyuki and Kameda, Daisuke and Suzuki, Hiroshi and Fukuda, Naoki and Takeda, Hiroyuki and Yanagisawa, Yoshiyuki and Ohtake, Masao and Kusaka, Kensuke and Yoshida, Koichi and Inabe, Naohito and Ohnishi, Tetsuya and Yoshida, Atsushi and Tanaka, Kanenobu and Mizoi, Yutaka},
    title = "{BigRIPS separator and ZeroDegree spectrometer at RIKEN RI Beam Factory}",
    journal = {Progress of Theoretical and Experimental Physics},
    volume = {2012},
    number = {1},
    year = {2012},
    month = {12},
    issn = {2050-3911},
    doi = {10.1093/ptep/pts064},
    url = {https://doi.org/10.1093/ptep/pts064},
    eprint = {http://oup.prod.sis.lan/ptep/article-pdf/2012/1/03C003/11595011/pts064.pdf},
}

@article{Fukuda2013,
title = "Identification and separation of radioactive isotope beams by the BigRIPS separator at the RIKEN RI Beam Factory",
journal = "Nuclear Instruments and Methods in Physics Research Section B: Beam Interactions with Materials and Atoms",
volume = "317",
pages = "323 - 332",
year = "2013",
note = "XVIth International Conference on ElectroMagnetic Isotope Separators and Techniques Related to their Applications, December 2–7, 2012 at Matsue, Japan",
issn = "0168-583X",
doi = "https://doi.org/10.1016/j.nimb.2013.08.048",
url = "http://www.sciencedirect.com/science/article/pii/S0168583X13009890",
author = "N. Fukuda and T. Kubo and T. Ohnishi and N. Inabe and H. Takeda and D. Kameda and H. Suzuki"
}

@article{Louchart2014,
title = "The PRESPEC liquid-hydrogen target for in-beam gamma spectroscopy of exotic nuclei at GSI",
journal = "Nuclear Instruments and Methods in Physics Research Section A: Accelerators, Spectrometers, Detectors and Associated Equipment",
volume = "736",
pages = "81 - 87",
year = "2014",
issn = "0168-9002",
doi = "https://doi.org/10.1016/j.nima.2013.10.035",
url = "http://www.sciencedirect.com/science/article/pii/S0168900213014150",
author = "C. Louchart and J.M. Gheller and Ph. Chesny and G. Authelet and J.Y. Rousse and A. Obertelli and P. Boutachkov and S. Pietri and F. Ameil and L. Audirac and A. Corsi and Z. Dombradi and J. Gerl and A. Gillibert and W. Korten and C. Mailleret and E. Merchan and C. Nociforo and N. Pietralla and D. Ralet and M. Reese and V. Stepanov"
}

@Article{Obertelli2014,
author="Obertelli, A.
and Delbart, A.
and Anvar, S.
and Audirac, L.
and Authelet, G.
and Baba, H.
and Bruyneel, B.
and Calvet, D.
and Ch{\^a}teau, F.
and Corsi, A.
and Doornenbal, P.
and Gheller, J. -M.
and Giganon, A.
and Lahonde-Hamdoun, C.
and Leboeuf, D.
and Loiseau, D.
and Mohamed, A.
and Mols, J. -Ph.
and Otsu, H.
and P{\'e}ron, C.
and Peyaud, A.
and Pollacco, E. C.
and Prono, G.
and Rousse, J. -Y.
and Santamaria, C.
and Uesaka, T.",
title="MINOS: A vertex tracker coupled to a thick liquid-hydrogen target for in-beam spectroscopy of exotic nuclei",
journal="The European Physical Journal A",
year="2014",
month="Jan",
day="30",
volume="50",
number="1",
pages="8",
issn="1434-601X",
doi="10.1140/epja/i2014-14008-y",
url="https://doi.org/10.1140/epja/i2014-14008-y"
}

@article{Kobayashi2013,
title = "SAMURAI spectrometer for RI beam experiments",
//journal = "Nuclear Instruments and Methods in Physics Research Section B: Beam Interactions with Materials and Atoms",
journal = "Nucl. Instrum. Methods Phys. Res., Sect. B",
volume = "317",
pages = "294 - 304",
year = "2013",
//note = "XVIth International Conference on ElectroMagnetic Isotope Separators and Techniques Related to their Applications, December 2–7, 2012 at Matsue, Japan",
issn = "0168-583X",
doi = "https://doi.org/10.1016/j.nimb.2013.05.089",
url = "http://www.sciencedirect.com/science/article/pii/S0168583X13007118",
author = "T. Kobayashi and N. Chiga and T. Isobe and Y. Kondo and T. Kubo and K. Kusaka and T. Motobayashi and T. Nakamura and J. Ohnishi and H. Okuno and H. Otsu and T. Sako and H. Sato and Y. Shimizu and K. Sekiguchi and K. Takahashi and R. Tanaka and K. Yoneda",
keywords = "RI beam, Large acceptance spectrometer, Invariant mass spectroscopy"
}

@article{Kondo2020,
title = {Recent progress and developments for experimental studies with the SAMURAI spectrometer},
journal = {Nuclear Instruments and Methods in Physics Research Section B: Beam Interactions with Materials and Atoms},
volume = {463},
pages = {173-178},
year = {2020},
issn = {0168-583X},
doi = {https://doi.org/10.1016/j.nimb.2019.05.068},
url = {https://www.sciencedirect.com/science/article/pii/S0168583X19303891},
author = {Y. Kondo and T. Tomai and T. Nakamura}
}

@article{Kubota2020,
  title = {Surface Localization of the Dineutron in $^{11}\mathrm{Li}$},
  author = {Kubota, Y. and Corsi, A. and Authelet, G. and Baba, H. and Caesar, C. and Calvet, D. and Delbart, A. and Dozono, M. and Feng, J. and Flavigny, F. and Gheller, J.-M. and Gibelin, J. and Giganon, A. and Gillibert, A. and Hasegawa, K. and Isobe, T. and Kanaya, Y. and Kawakami, S. and Kim, D. and Kikuchi, Y. and Kiyokawa, Y. and Kobayashi, M. and Kobayashi, N. and Kobayashi, T. and Kondo, Y. and Korkulu, Z. and Koyama, S. and Lapoux, V. and Maeda, Y. and Marqu\'es, F. M. and Motobayashi, T. and Miyazaki, T. and Nakamura, T. and Nakatsuka, N. and Nishio, Y. and Obertelli, A. and Ogata, K. and Ohkura, A. and Orr, N. A. and Ota, S. and Otsu, H. and Ozaki, T. and Panin, V. and Paschalis, S. and Pollacco, E. C. and Reichert, S. and Rouss\'e, J.-Y. and Saito, A. T. and Sakaguchi, S. and Sako, M. and Santamaria, C. and Sasano, M. and Sato, H. and Shikata, M. and Shimizu, Y. and Shindo, Y. and Stuhl, L. and Sumikama, T. and Sun, Y. L. and Tabata, M. and Togano, Y. and Tsubota, J. and Yang, Z. H. and Yasuda, J. and Yoneda, K. and Zenihiro, J. and Uesaka, T.},
  journal = {Phys. Rev. Lett.},
  volume = {125},
  issue = {25},
  pages = {252501},
  numpages = {7},
  year = {2020},
  month = {Dec},
  publisher = {American Physical Society},
  doi = {10.1103/PhysRevLett.125.252501},
  url = {https://link.aps.org/doi/10.1103/PhysRevLett.125.252501}
}

@article{Corsi2019,
title = {Structure of 13Be probed via quasi-free scattering},
journal = {Physics Letters B},
volume = {797},
pages = {134843},
year = {2019},
issn = {0370-2693},
doi = {https://doi.org/10.1016/j.physletb.2019.134843},
url = {https://www.sciencedirect.com/science/article/pii/S037026931930557X},
author = {A. Corsi and Y. Kubota and J. Casal and M. Gómez-Ramos and A.M. Moro and G. Authelet and H. Baba and C. Caesar and D. Calvet and A. Delbart and M. Dozono and J. Feng and F. Flavigny and J.-M. Gheller and J. Gibelin and A. Giganon and A. Gillibert and K. Hasegawa and T. Isobe and Y. Kanaya and S. Kawakami and D. Kim and Y. Kiyokawa and M. Kobayashi and N. Kobayashi and T. Kobayashi and Y. Kondo and Z. Korkulu and S. Koyama and V. Lapoux and Y. Maeda and F.M. Marqués and T. Motobayashi and T. Miyazaki and T. Nakamura and N. Nakatsuka and Y. Nishio and A. Obertelli and A. Ohkura and N.A. Orr and S. Ota and H. Otsu and T. Ozaki and V. Panin and S. Paschalis and E.C. Pollacco and S. Reichert and J.-Y. Rousse and A.T. Saito and S. Sakaguchi and M. Sako and C. Santamaria and M. Sasano and H. Sato and M. Shikata and Y. Shimizu and Y. Shindo and L. Stuhl and T. Sumikama and Y.L. Sun and M. Tabata and Y. Togano and J. Tsubota and T. Uesaka and Z.H. Yang and J. Yasuda and K. Yoneda and J. Zenihiro}
}

@article{Yang2021,
  title = {Quasifree Neutron Knockout Reaction Reveals a Small $s$-Orbital Component in the Borromean Nucleus $^{17}\mathrm{B}$},
  author = {Yang, Z. H. and Kubota, Y. and Corsi, A. and Yoshida, K. and Sun, X.-X. and Li, J. G. and Kimura, M. and Michel, N. and Ogata, K. and Yuan, C. X. and Yuan, Q. and Authelet, G. and Baba, H. and Caesar, C. and Calvet, D. and Delbart, A. and Dozono, M. and Feng, J. and Flavigny, F. and Gheller, J.-M. and Gibelin, J. and Giganon, A. and Gillibert, A. and Hasegawa, K. and Isobe, T. and Kanaya, Y. and Kawakami, S. and Kim, D. and Kiyokawa, Y. and Kobayashi, M. and Kobayashi, N. and Kobayashi, T. and Kondo, Y. and Korkulu, Z. and Koyama, S. and Lapoux, V. and Maeda, Y. and Marqu\'es, F. M. and Motobayashi, T. and Miyazaki, T. and Nakamura, T. and Nakatsuka, N. and Nishio, Y. and Obertelli, A. and Ohkura, A. and Orr, N. A. and Ota, S. and Otsu, H. and Ozaki, T. and Panin, V. and Paschalis, S. and Pollacco, E. C. and Reichert, S. and Rouss\'e, J.-Y. and Saito, A. T. and Sakaguchi, S. and Sako, M. and Santamaria, C. and Sasano, M. and Sato, H. and Shikata, M. and Shimizu, Y. and Shindo, Y. and Stuhl, L. and Sumikama, T. and Sun, Y. L. and Tabata, M. and Togano, Y. and Tsubota, J. and Xu, F. R. and Yasuda, J. and Yoneda, K. and Zenihiro, J. and Zhou, S.-G. and Zuo, W. and Uesaka, T.},
  journal = {Phys. Rev. Lett.},
  volume = {126},
  issue = {8},
  pages = {082501},
  numpages = {8},
  year = {2021},
  month = {Feb},
  publisher = {American Physical Society},
  doi = {10.1103/PhysRevLett.126.082501},
  url = {https://link.aps.org/doi/10.1103/PhysRevLett.126.082501}
}

@article{Fossez2018,
  title = {Energy spectrum of neutron-rich helium isotopes: Complex made simple},
  author = {Fossez, K. and Rotureau, J. and Nazarewicz, W.},
  journal = {Phys. Rev. C},
  volume = {98},
  issue = {6},
  pages = {061302},
  numpages = {6},
  year = {2018},
  month = {Dec},
  publisher = {American Physical Society},
  doi = {10.1103/PhysRevC.98.061302},
  url = {https://link.aps.org/doi/10.1103/PhysRevC.98.061302}
}

@article{Bohlen1999,
title = {Spectroscopy of exotic nuclei with multi-nucleon transfer reactions},
journal = {Progress in Particle and Nuclear Physics},
volume = {42},
pages = {17-26},
year = {1999},
note = {Heavy Ion Collisions from Nuclear to Quark Matter},
issn = {0146-6410},
doi = {https://doi.org/10.1016/S0146-6410(99)00056-3},
url = {https://www.sciencedirect.com/science/article/pii/S0146641099000563},
author = {H.G. Bohlen and A. Blazevic and B. Gebauer and W. {Von Oertzen} and S. Thummerer and R. Kalpakchieva and S.M. Grimes and T.N. Massey}
}

@article{Oertzen1995,
title = {Nuclear structure studies of very neutron-rich isotopes of 7–10He, 9–11Li and 12–14Be via two-body reactions},
journal = {Nuclear Physics A},
volume = {588},
number = {1},
pages = {c129-c134},
year = {1995},
note = {Proceedings of the Fifth International Symposium on Physics of Unstable Nuclei},
issn = {0375-9474},
doi = {https://doi.org/10.1016/0375-9474(95)00111-D},
url = {https://www.sciencedirect.com/science/article/pii/037594749500111D},
author = {W. {von Oertzen} and H.G. Bohlen and B. Gebauer and M. {von Lucke-Petsch} and A.N. Ostrowski and Ch. Seyfert and Th. Stolla and M. Wilpert and Th. Wilpert and D.V. Alexandrov and A.A. Korsheninnikov and I. Mukha and A.A. Ogloblin and R. Kalpakchieva and Y.E. Penionzhkevich and S. Piskor and S.M. Grimes and T.N. Massey}
}

@article{Rogachev2003,
  title = {$T=5/2$ states in ${}^{9}\mathrm{Li}:$ Isobaric analog states of ${}^{9}\mathrm{He}$},
  author = {Rogachev, G. V. and Goldberg, V. Z. and Kolata, J. J. and Chubarian, G. and Aleksandrov, D. and Fomichev, A. and Golovkov, M. S. and Oganessian, Yu. Ts. and Rodin, A. and Skorodumov, B. and Slepnev, R. S. and Ter-Akopian, G. and Trzaska, W. H. and Wolski, R.},
  journal = {Phys. Rev. C},
  volume = {67},
  issue = {4},
  pages = {041603},
  numpages = {5},
  year = {2003},
  month = {Apr},
  publisher = {American Physical Society},
  doi = {10.1103/PhysRevC.67.041603},
  url = {https://link.aps.org/doi/10.1103/PhysRevC.67.041603}
}

@article{Uberseder2016,
title = {Nuclear structure beyond the neutron drip line: The lowest energy states in 9He via their T=5/2 isobaric analogs in 9Li},
journal = {Physics Letters B},
volume = {754},
pages = {323-327},
year = {2016},
issn = {0370-2693},
doi = {https://doi.org/10.1016/j.physletb.2016.01.014},
url = {https://www.sciencedirect.com/science/article/pii/S0370269316000186},
author = {E. Uberseder and G.V. Rogachev and V.Z. Goldberg and E. Koshchiy and B.T. Roeder and M. Alcorta and G. Chubarian and B. Davids and C. Fu and J. Hooker and H. Jayatissa and D. Melconian and R.E. Tribble}
}

@article{Vorabbi2018,
  title = {Structure of the exotic $^{9}\mathrm{He}$ nucleus from the no-core shell model with continuum},
  author = {Vorabbi, Matteo and Calci, Angelo and Navr\'atil, Petr and Kruse, Michael K. G. and Quaglioni, Sofia and Hupin, Guillaume},
  journal = {Phys. Rev. C},
  volume = {97},
  issue = {3},
  pages = {034314},
  numpages = {9},
  year = {2018},
  month = {Mar},
  publisher = {American Physical Society},
  doi = {10.1103/PhysRevC.97.034314},
  url = {https://link.aps.org/doi/10.1103/PhysRevC.97.034314}
}

@article{Kalanee2013,
  title = {Structure of unbound neutron-rich ${}^{9}$He studied using single-neutron transfer},
  author = {Al Kalanee, T. and Gibelin, J. and Roussel-Chomaz, P. and Keeley, N. and Beaumel, D. and Blumenfeld, Y. and Fern\'andez-Dom\'{\i}nguez, B. and Force, C. and Gaudefroy, L. and Gillibert, A. and Guillot, J. and Iwasaki, H. and Krupko, S. and Lapoux, V. and Mittig, W. and Mougeot, X. and Nalpas, L. and Pollacco, E. and Rusek, K. and Roger, T. and Savajols, H. and de S\'er\'eville, N. and Sidorchuk, S. and Suzuki, D. and Strojek, I. and Orr, N. A.},
  journal = {Phys. Rev. C},
  volume = {88},
  issue = {3},
  pages = {034301},
  numpages = {8},
  year = {2013},
  month = {Sep},
  publisher = {American Physical Society},
  doi = {10.1103/PhysRevC.88.034301},
  url = {https://link.aps.org/doi/10.1103/PhysRevC.88.034301}
}

@article{Nollett2012,
  title = {Ab initio calculations of nuclear widths via an integral relation},
  author = {Nollett, Kenneth M.},
  journal = {Phys. Rev. C},
  volume = {86},
  issue = {4},
  pages = {044330},
  numpages = {20},
  year = {2012},
  month = {Oct},
  publisher = {American Physical Society},
  doi = {10.1103/PhysRevC.86.044330},
  url = {https://link.aps.org/doi/10.1103/PhysRevC.86.044330}
}

@article{Volya2005,
  title = {Discrete and Continuum Spectra in the Unified Shell Model Approach},
  author = {Volya, Alexander and Zelevinsky, Vladimir},
  journal = {Phys. Rev. Lett.},
  volume = {94},
  issue = {5},
  pages = {052501},
  numpages = {4},
  year = {2005},
  month = {Feb},
  publisher = {American Physical Society},
  doi = {10.1103/PhysRevLett.94.052501},
  url = {https://link.aps.org/doi/10.1103/PhysRevLett.94.052501}
}

@article{Barker2004,
title = {Level widths in 9He and 10He},
journal = {Nuclear Physics A},
volume = {741},
pages = {42-51},
year = {2004},
issn = {0375-9474},
doi = {https://doi.org/10.1016/j.nuclphysa.2004.06.001},
url = {https://www.sciencedirect.com/science/article/pii/S0375947404007389},
author = {F.C Barker}
}

@misc{Rogachev_private,
   author = "G. V. Rogachev",
   howpublished = "private communication"
}

\end{document}